\documentclass[aps,prd,preprint,groupedaddress,showkeys,showpacs]{revtex4-2}
\usepackage{graphicx,epsfig,dcolumn,bm,epic,eepic,float}
\usepackage{amsmath}
\usepackage{hyperref}
\usepackage{latexsym}
\usepackage[noabbrev]{cleveref}
\usepackage{color}
\usepackage{cancel}
\usepackage{makeidx,shortvrb,latexsym}
\begin{document}
\unitlength 1 cm
\newcommand{\nn}{\nonumber}
\newcommand{\vk}{\vec k}
\newcommand{\vp}{\vec p}
\newcommand{\vq}{\vec q}
\newcommand{\vkp}{\vec {k'}}
\newcommand{\vpp}{\vec {p'}}
\newcommand{\vqp}{\vec {q'}}
\newcommand{\bk}{{\bf k}}
\newcommand{\bp}{{\bf p}}
\newcommand{\bq}{{\bf q}}
\newcommand{\br}{{\bf r}}
\newcommand{\bR}{{\bf R}}
\newcommand{\up}{\uparrow}
\newcommand{\down}{\downarrow}
\newcommand{\fns}{\footnotesize}
\newcommand{\ns}{\normalsize}
\newcommand{\cdag}{c^{\dagger}}

\title {On the ambiguity between   differential and integral forms of the Martin-Ryskin-Watt unintegrated parton distribution function model}
\author{\textit{Ramin Kord Valeshabadi}}%
\author{\textit{M. Modarres} }
\email[Corresponding author: ]{mmodares@ut.ac.ir}
\affiliation {Department of Physics, University of $Tehran$,
	1439955961, $Tehran$, Iran.}
\date{\today}
\begin{abstract}
In this work, we study the structure of the leading order Martin-Ryskin-Watt (MRW) unintegrated parton distribution function (UPDF) and explain in detail why there exists discrepancy between the two different definitions of this UPDF model, i.e., the integral (I-MRW) and differential (D-MRW) MRW UPDFs. We perform this investigation with both angular and strong ordering cutoffs. The derivation footsteps of obtaining the I-MRW UPDF   from the D-MRW ones are numerically performed, and the reason of such non-equivalency between the two forms is clearly explained.  We show and find out that both   methods suggested in the papers by  Golec-Biernat and Staśto as well as that of  Guiot  have shortcomings, and only the combination of their prescriptions can give us the same UPDF structure from both of these two different versions of the MRW UPDF, namely  I-MRW  and  the D-MRW UPDFs.
\end{abstract}
 \maketitle

\section{Introduction}
\label{sec:I}
Unintegrated parton distribution functions (UPDFs) are one of the essential ingredients of theoretical   hadronic cross sections calculation, within the $k_t$-factorization scheme ($k_t$ is the transverse momentum of a parton). In contrast to the collinear factorization framework, where partons evolve according to the DGLAP evolution equations, the evolution equations within the $k_t$-factorization is only limited to the gluon, i.e., Balitsky–Fadin–Kuraev–
Lipatov (BFLK) \cite{BFKL1,BFKL2,BFKL3,BFKL4} and  Catani-Ciafaloni-Fiorani-Marchesini (CCFM) \cite{CCFM1,CCFM2,CCFM3,CCFM4}. Therefore, different methods are introduced for obtaining both quarks and gluon UPDFs within the $k_t$-factorization framework, which are mostly based on the DGLAP evolution equations.
Among these methods  Kimber-Martin-Ryskin (KMR) \cite{KMR}, Martin-Ryskin-Watt (MRW) \cite{MRW}, and parton-branching (PB) \cite{PB1,PB2} are mostly used in the phenomenological  study and successfully they could describe the experimental data \cite{Mod16,PB_Z_production, kord_inclusive_jet,Valeshabadi:2021twn,kord_validity,taghavi_drell}. In the MRW formalism, which is the main focus of this work, it is assumed that the parton moves collinear to the incoming  proton till the last evolution step, where it becomes $k_t$ dependent, and after emitting a real emission evolves to the factorization scales with the help of Sudakov form factor. While in the PB UPDFs,  the $k_t$ dependency enters into formalism from the beginning of the evolution via an initial Gaussian distribution. Then the UPDFs are obtained by using  Monte Carlo (PB) method and taking into account the transverse momentum of the parton along the evolution ladder. 

But, the MRW formalism at the leading order (LO) level, can be written in two alternative forms, i.e., integral (I-MRW) and differential  (D-MRW) UPDFs derivations. The apparent equivalency of these two forms becomes   questionable in the reference \cite{Golec_on_KMR}, where it is shown that these two versions can actually become different in certain regions of $x$ (x is the fractional momentum) and $k_t$. In order to address this problem, the authors of   reference \cite{Golec_on_KMR} suggested that only in the cutoff-dependent parton distribution functions (PDFs) can solve this discrepancy. On the other hand, the    reference \cite{Guiot_pathology} contradicts the above idea \cite{Golec_on_KMR} and claims that there is no need for the cutoff-dependent PDFs, if one introduces  another term   to the D-MRW UPDF. 

However, in this work we show that both of the solutions suggested in the references \cite{Golec_on_KMR,Guiot_pathology} are incomplete, and the true equality between the I-MRW and D-MRW UPDFs derivations can only be obtained if the cutoff-dependent PDFs and the additional term at the same time be included into the formalism.

The structure of the paper is as follows: In the \cref{sec:1}, the integral and differential forms of the MRW UPDFs are in detail explained. In \cref{sec:2}, we show the numerical results of I-MRW and D-MRW UPDFs, to explain why one obtains different results, and also what should be done in order to bring back the equivalency between the two forms. In the \cref{sec:3}, we derive the same UPDFs from both of the I-MRW and D-MRW UPDFs, using our analysis in the \cref{sec:2}. Finally, in the \cref{sec:4},  the conclusions are presented.
	
\section{Integral and differential forms of the MRW UPDFs}
\label{sec:1}
The MRW model as explained in the introduction can simply be obtained by assuming the evolution of parton collinear to the parent hadron, till the last evolution step. At this step, the  parton in the last evolution step becomes $k_t$ dependent, i.e. $f_b(x, k_t^2)$. Then the parton emits a real emission with the probability $\dfrac{\alpha_s(k_t^2)}{2 \pi} P_{ab}(x/z)$, and finally evolves to the factorization scale $\mu$ with the help of the Sudakov form factor $T_a(k_t^2, \mu^2)$, i.e.,:
\begin{equation}
	\label{eq:1}
	f_a(x, k_t^2, \mu^2) = T_a(k_t^2, \mu^2) \dfrac{\alpha_s(k_t^2)}{2 \pi k_t^2} \sum_{b=q,g} \int_x^1 dz P_{ab}(z) f_b(\dfrac{x}{z}, k_t^2 )[\Theta(z_{max} - z)]^{\delta^{ab}},
\end{equation}
where the Sudakov form factor is as follows:
\begin{equation}
	\label{eq:2}
	T_a(k_t^2, \mu^2) = exp\left(-\int_{k_t^2}^{\mu^2}  \dfrac{dk_t^2}{k_t^2} \dfrac{\alpha_s(k_t^2)}{2 \pi} \sum_{b=q,g} \int_0^1 d\xi \xi P_{ba}(\xi)[\Theta(\xi_{max} - \xi)]^{\delta^{ab}} \right) ,
\end{equation}
with
\begin{equation}
	\label{eq:3}
	T_a(k_t^2 > \mu^2, \mu^2) = 1.
\end{equation}
It should be noted that in the \cref{eq:1} the momentum weighted PDFs are used, i.e. $f_b(x, k_t^2)=x b(x, k_t^2)$. Because of using  the collinear input PDFs with this approach, the MRW formalism is not valid at $k_t$ less than a certain starting point, $\mu_0 \sim 1\; GeV$. Therefore, to define the UPDFs at $k_t < \mu_0$, one can utilize the normalization condition as a constraint, i.e.,:
\begin{equation}
	f(x, \mu^2) = \int_{0}^{\mu^2} dk_t^2 f(x, k_t^2, \mu^2),
\end{equation}
and obtain UPDFs at $k_t < \mu_0$ as \cite{MRW}:
\begin{equation}
	f_a(x, k_t^2 < \mu_0^2, \mu^2) = T_a(\mu_0^2, \mu^2) f(x, \mu_0^2).
\end{equation}
Expanding the \cref{eq:1,eq:2} for the quark and gluon, gives divergent behavior for the probability terms corresponding to the soft gluon emission, i.e., $P_{qq}$ and $P_{gg}$. However, the Heaviside step function avoids this soft gluon divergences. On the other hand we should note that in the Kimber-Martin-Ryskin (KMR) model \cite{KMR} which is used by the reference \cite{Golec_on_KMR}, this cutoff is wrongly imposed on both emissions.    

In the literature two kinds of cutoffs are used. The first one that is most commonly used is based on the angular ordering constraint (AOC) of the soft gluon emissions. Imposing this constraint on the last emission step, i.e., $z \tilde{q}_t= z \dfrac{k_t}{(1-z)} < \mu$ where $\tilde{q}_t$ is the rescale transverse momentum \cite{MRW, Valeshabadi:2021twn, Golec_on_KMR}, leads to  the cutoff on $z$ which can be obtained as follows:
\begin{equation}
	\label{eq:4}
	z_{max} = \dfrac{\mu}{(\mu + k_t)}.
\end{equation}
Using the above cutoff allows the parton to have emission even at the $k_t > \mu$, and hence the UPDFs can become large in this limit, mostly because the Sudakov form factor is limited to the $k_t < \mu$. The other cutoff on $z$ can be obtained by using the strong ordering constraint (SOC) of the gluon emission, i.e., $\tilde{q}_t= \dfrac{k_t}{(1-z)} < \mu$ \cite{KimberSO}:
\begin{equation}
	\label{eq:5}
	z_{max} = 1 - \dfrac{k_t}{\mu}.
\end{equation}
The SOC is harsher with respect to the AOC one, and it limits the transverse momentum of emitted gluon to the $k_t < \mu$. However, we should be aware that parton within the MRW UPDF model is still free to have transverse momentum larger than the factorization scale via the quark emission term. 

Although, in the MRW model, the parton has the freedom to have the transverse momentum larger than the factorization scale via the quark emission term, but within the KMR model, the parton is limited to the $k_t \leq \mu$, due to the cutoff on both emission terms. As a result of this, one can notice from the figure $1$ of the reference \cite{Golec_on_KMR}  that the UPDF model adopted in this reference is in fact the KMR model, in which the author  of reference \cite{Guiot_pathology} not correctly refers to it as the MRW UPDF. Also, we should point out that the same author uses the strong ordering cutoff along with the hard constraint $\Theta(\mu-k_t)$, to limit the parton transverse momentum to the $k_t \le \mu$. 

From now on, in order to simply   prove the above points and show  how to remove this discrepancies, in the following sections, we  only consider the non-singlet (NS) quark   UPDF, i.e.:
\begin{equation}
	\label{eq:6}
	f_q^{NS}(x, k_t^2, \mu^2) = T_q^{NS}(k_t^2, \mu^2) \dfrac{\alpha_s(k_t^2)}{2 \pi k_t^2} \int_x^{z_{max}} dz P_{qq}(z) f_q^{NS}(\dfrac{x}{z}, k_t^2),
\end{equation} 
where $f_q^{NS}(x, k_t^2) = \sum_{i=1}^{Nf} (f_i(x, k_t^2) -\overline{f}_i(x, k_t^2)) $ is the non-singlet distribution, and the Sudakov form factor for this distribution is:
\begin{equation}
	\label{eq:7}
	T_q^{NS}(k_t^2, \mu^2) = exp\left(-\int_{k_t^2}^{\mu^2}  \dfrac{dk_t^2}{k_t^2} \dfrac{\alpha_s(k_t^2)}{2 \pi} \int_0^{z_{max}} d\xi \xi P_{qq}(\xi) \right).
\end{equation}

We should note that in the non-singlet distribution the KMR and MRW UPDFs have the same form, due to this fact that the non-diagonal quark emission terms of the DGLAP evolution equation are  no longer   exist. The most important benefit of using this distribution is that the DGLAP evolution equation is not a coupled integro-differential evolution any more, and we can simply obtain the cutoff-dependent PDFs at different $k_t^2$, see the \cref{sec:2}.    

The MRW UPDF   explained above is usually written in its integral form, I-MRW. However, in the reference \cite{MRW} it is shown that it can also be written as a compact D-MRW UPDF   as follows:
\begin{equation}
	\label{eq:8}
	f_q^{NS}(x, k_t^2, \mu^2) = \dfrac{\partial}{\partial k_t^2}[f_q^{NS}(x, k_t^2) T_q^{NS}(k_t^2, \mu^2)] = T_q^{NS}(k_t^2, \mu^2) \dfrac{\alpha_s(k_t^2)}{2 \pi k_t^2} \int_x^1 dz P_{qq}(z) f_q^{NS}(\dfrac{x}{z}, k_t^2).
\end{equation}
In order to reach from the D-MRW to the I-MRW one can do as follows:
\begin{equation}
	\label{eq:9}
	\dfrac{\partial}{\partial k_t^2}[f_q^{NS}(x, k_t^2) T_q^{NS}(k_t^2, \mu^2)] = T_q^{NS}(k_t^2, \mu^2) \dfrac{\partial f_q^{NS}(x, k_t^2) }{\partial k_t^2} + f_q^{NS}(x, k_t^2) \dfrac{d T_q^{NS}(k_t^2, \mu^2) }{d k_t^2},
\end{equation}
now the derivative with respect to $ k_t^2$ can be written as the form of the modified DGLAP evolution equation (MDGLAP), i.e.:
\begin{equation}
	\label{eq:10}
	\dfrac{\partial f_q^{NS}(x, k_t^2) }{\partial k_t^2} = \dfrac{\alpha_s(k_t^2)}{2 \pi k_t^2}\left[ \int_x^{z_{max}} P_{qq}(z) f_q^{NS}(\dfrac{x}{z}, k_t^2) - f_q^{NS}(x, k_t^2) \int_0^{z_{Max}} z P_{qq}(z) \right],
\end{equation} 
and using the following relation for the Sudakov form factor:
\begin{equation}
	\label{eq:11}
	\dfrac{1}{T_q^{NS}(k_t^2, \mu^2)} \dfrac{\partial T_q^{NS}(k_t^2, \mu^2) }{\partial k_t^2} = \dfrac{\alpha_s(k_t^2)}{2 \pi k_t^2} \int_0^{z_{max}} z P_{qq}(z),
\end{equation}
one can simply obtain the  I-MRW UPDF. The important point about this derivation is that the derivative with respect to the Sudakov form factor has the role to remove the virtual contribution, which comes from the modified MDGLAP. Considering this fact, therefore one expects that the I-MRW UPDF to be always positive.
\section{Numerical Investigation of the D-MRW and I-MRW UPDF}
\label{sec:2}
In this section we explain the  D-MRW and I-MRW UPDFs   by considering only the first three quarks NS distribution, i.e. $f_q^{NS}(x, k_t^2, \mu^2) = \sum_{q \in {u,d,s}} [ f_q(x, k_t^2, \mu^2) - f_{\overline{q}}(x, k_t^2, \mu^2)]$. For the calculation, we consider the central MSTW2008lo90cl-nf3 (MSTW) input PDF sets \cite{mstw} via the LHAPDF library \cite{LHAPDF6}. We calculate the UPDFs with respect to $k_t^2$ at different values of $x=0.01$ and $x=0.1$  with $\mu^2 = 100\; GeV^2$. 

\begin{figure}
	\includegraphics[width=8cm, height=8cm]{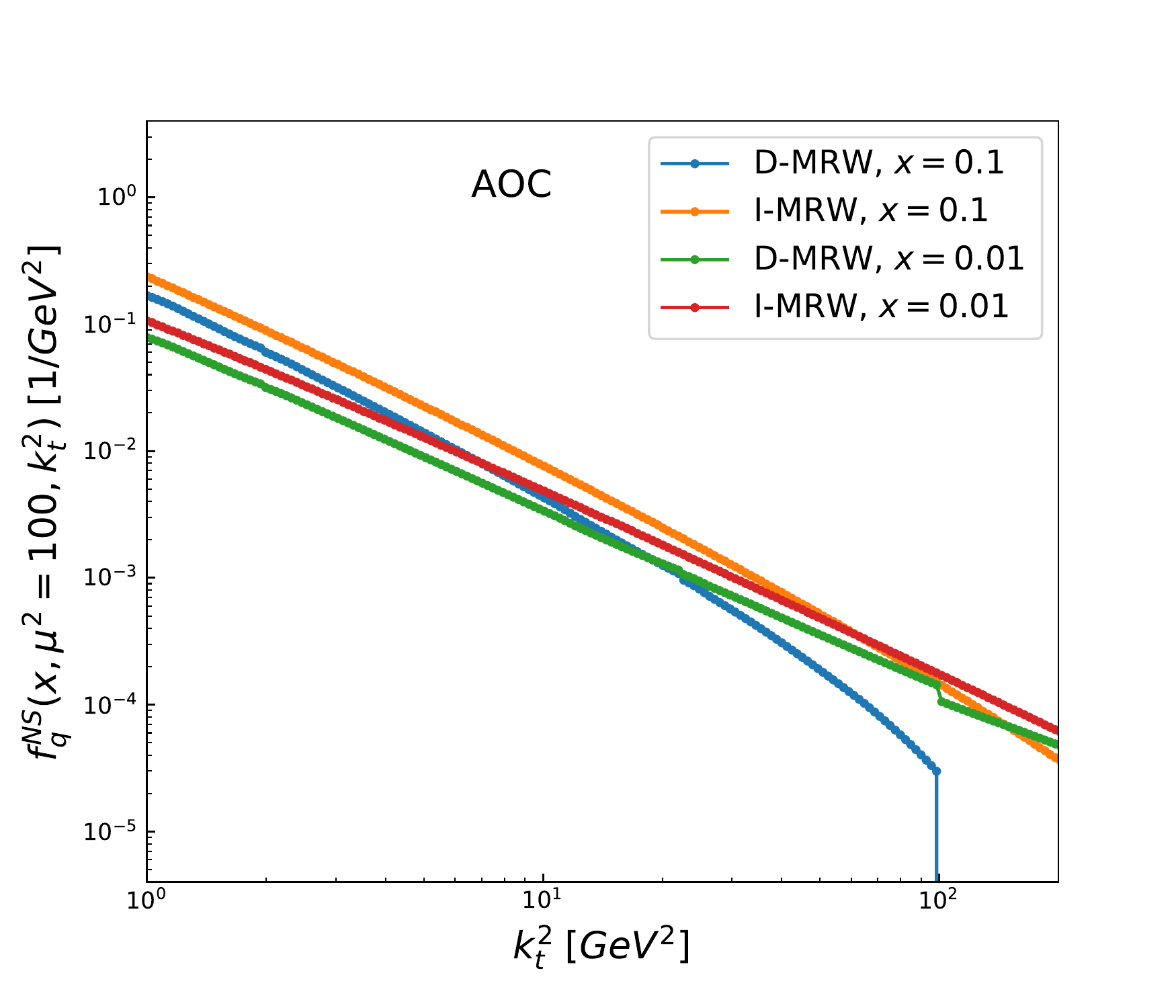}
	\includegraphics[width=8cm, height=8cm]{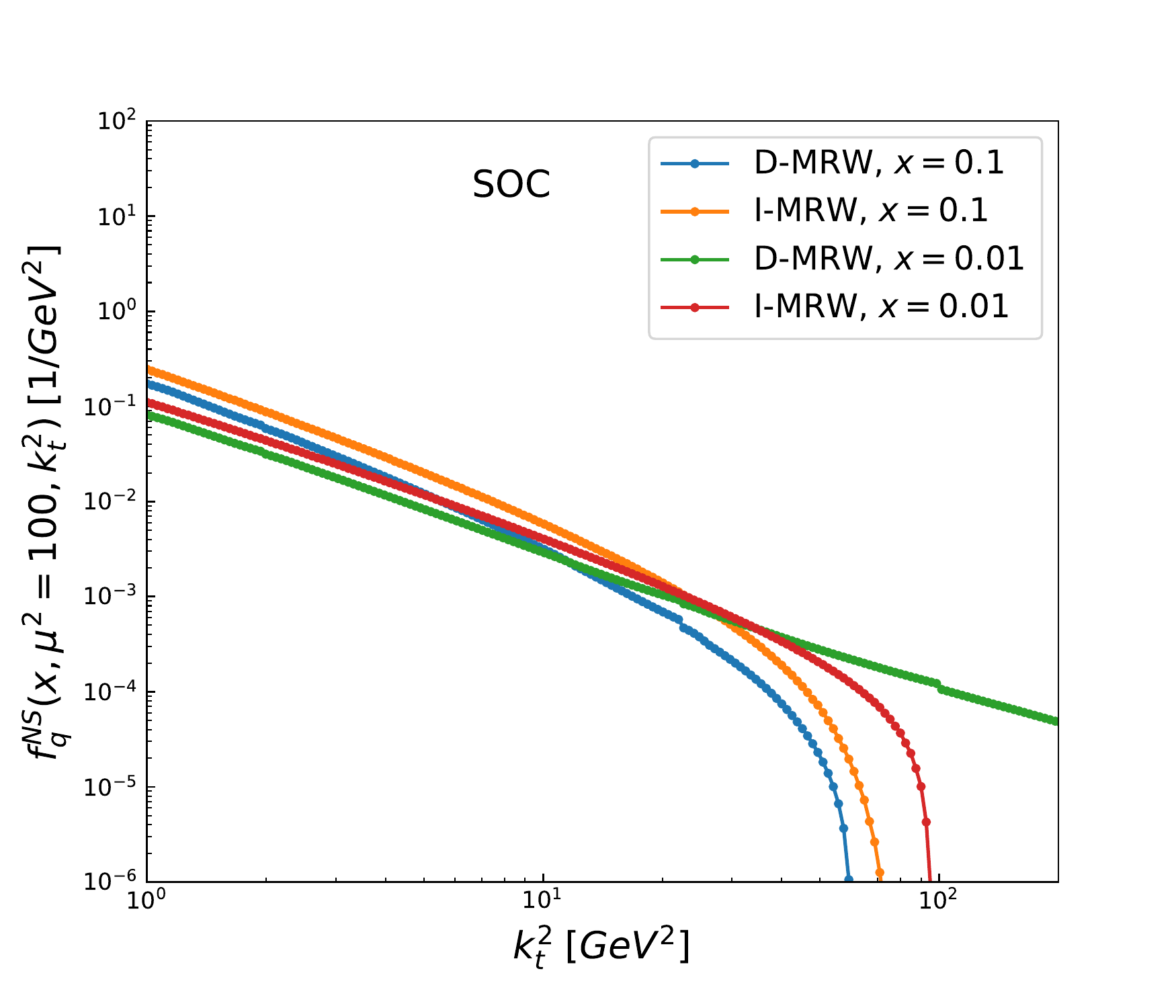}
	\caption
	{The left (right) panel shows the comparison of the non-singlet quark MRW UPDF with the AOC (SOC). The differential (integral) form of the \cref{eq:8} is denoted by D-MRW (I-MRW).
	}
	\label{fig:1}
\end{figure}

Looking at the \cref{fig:1}, it makes clear the issues related to the equality of the differential and integral forms of the MRW pointed out in the references \cite{Guiot_pathology,Golec_on_KMR}.   In the case of MRW with AOC,  two problems can be spotted quickly by looking at this figure. First, as we move toward the large $x$ limit, the difference between the two   forms is more significant, and at some points, even in the $k_t < \mu$ D-MRW UPDF becomes negative, while the I-MRW UPDF  is always positive. Second, the D-MRW  has a discontinuity at $k_t = \mu$. Also in the case of the I(D)-MRW UPDF with SOC, one can observe the same issues as the case of the I(D)-MRW UPDF with AOC, but, since  there is no  quark emission terms, the UPDFs with the integral form are suppressed down to zero. In order to understand the roots of these problems, we show numerically the derivation steps of reaching to the I-MRW UPDF from the D-MRW UPDF, i.e., the \cref{eq:10,eq:11}.

\begin{figure}
	\includegraphics[width=8cm, height=8cm]{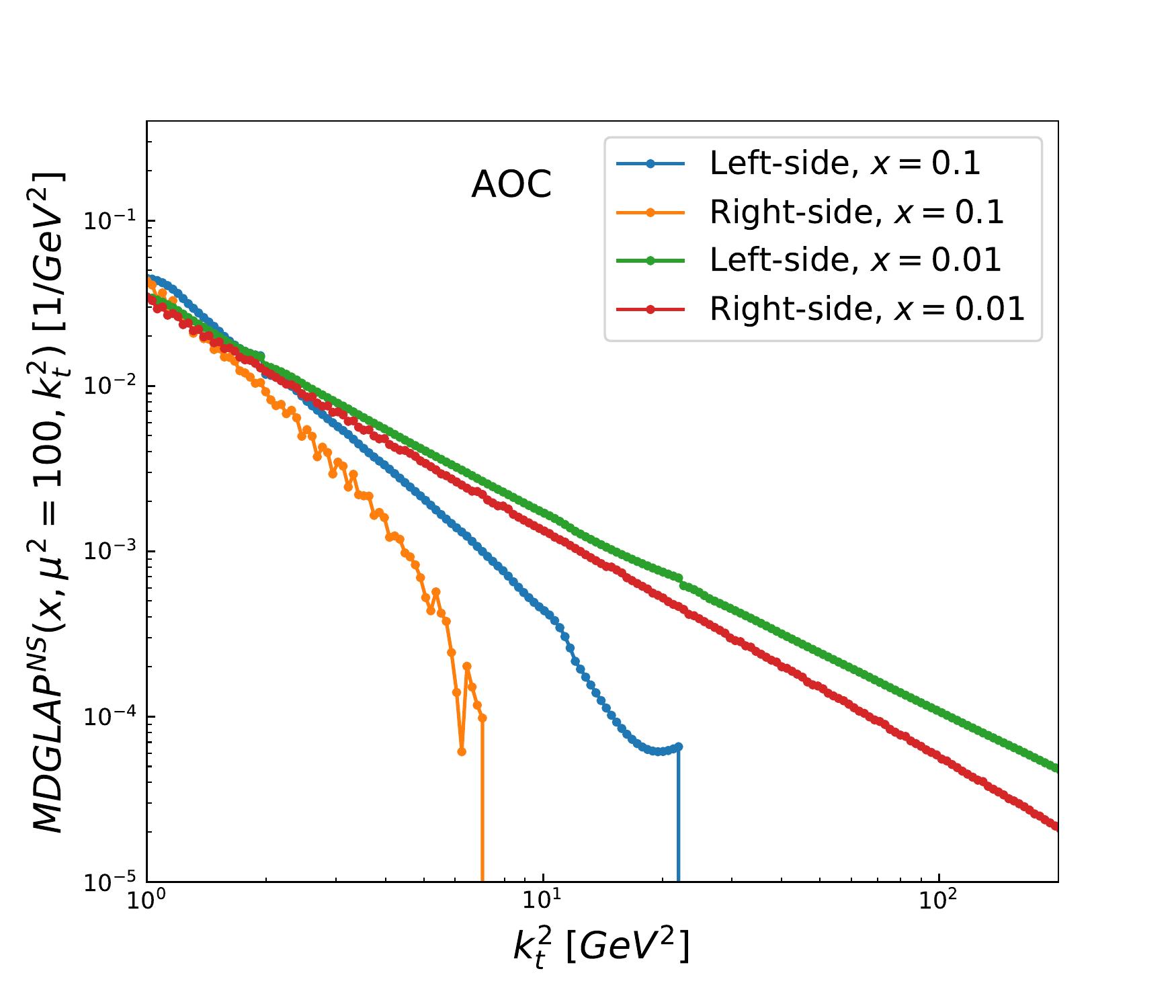}
	\includegraphics[width=8cm, height=8cm]{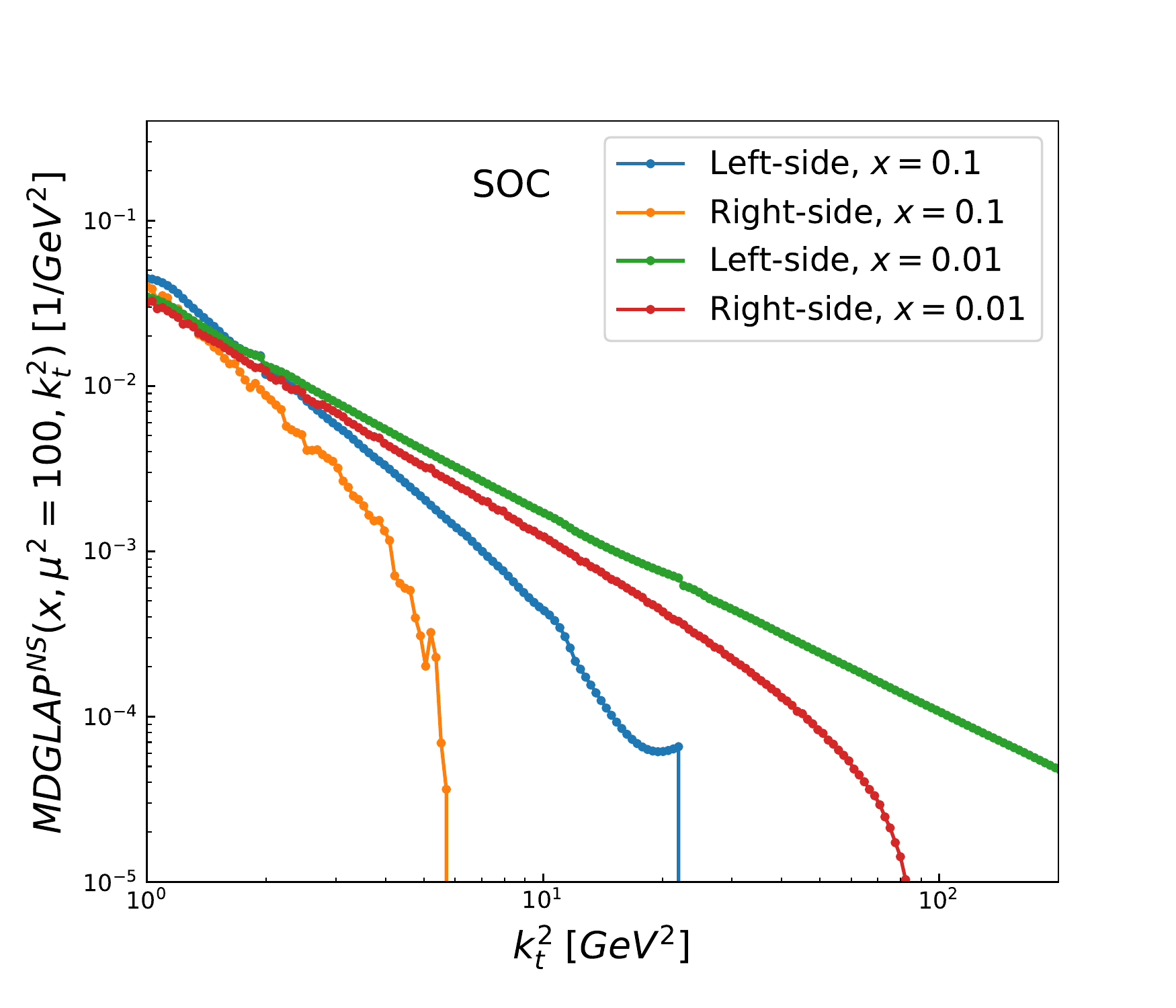}
	\caption
	{
		The left (right) panel shows the comparison of the left and right hand sides of the modified  MDGLAP for the non-singlet quark distribution with the AOC (SOC).
	}
	\label{fig:2}
\end{figure}

In the \cref{fig:2}, we   numerically demonstrate the validity of the left and right hand sides of the \cref{eq:10}, i.e., the MDGLAP with collinear PDFs. It can be seen from this figure that when the parton transverse momentum increases and becomes close to the factorization scale, or as $x$ becomes large, the difference between the left and right hand sides of the MDGLAP become more significant. This is actually related to the imposition of the soft gluon emission cutoff on the final evolution step in the right hand side of the MDGLAP. While in the left hand side, we only use the PDFs input, that has no cutoff on it. Henceforth, in order to solve this discrepancy, in the left hand side of the MDGLAP, one has to use PDFs with AOC and SOC imposed in the last evolution step. However, this is an arduous task, and one can alternatively solves the \cref{eq:10} with the cutoff on all evolution steps. In the reference \cite{Golec_on_KMR}, one needs  the  cutoff dependent-PDFs, in order to reach this equivalency between the  I-MRW and D-MRW UPDFs. However, as it is discussed in the following paragraphs and also in the reference \cite{Guiot_pathology}, it is  questionable, how the same UPDFs from  the I-MRW and D-MRW UPDFs in the $k_t \geq \mu$ is obtained. It should also be mentioned again that if one considers MRW quark distributions, and not the non-singlet one, then in the case of  SOC with the I-MRW UPDF, the UPDF has a tail at $k_t > \mu$. 

The problem with discontinuity and different distribution arising from the I-MRW and D-MRW UPDFs  is related to the this fact that \cref{eq:11} does not hold in the $k_t > \mu$. Because if we look carefully at the \cref{eq:7}, we   observe that the maximum value of the integral over $k_t^2$ is $\mu^2$, and as a result of this, the \cref{eq:11} is only truly valid at $k_t \leq \mu$. Henceforth, one should modify and correct this equation simply by adding the virtual term explicitly for $k_t > \mu$:
   \begin{equation}
   	\label{eq:12}
   	\dfrac{1}{T_q^{NS}(k_t^2, \mu^2)} \dfrac{\partial T_q^{NS}(k_t^2, \mu^2) }{\partial k_t^2} + \Theta(k_t^2 - \mu^2)\dfrac{\alpha_s(k_t^2)}{2 \pi k_t^2} \int_0^{z_{max}} z P_{qq}(z) = \dfrac{\alpha_s(k_t^2)}{2 \pi k_t^2} \int_0^{z_{max}} z P_{qq}(z).
   \end{equation}
As a result, one can modify the \cref{eq:8} as follows:
\begin{equation}
	\begin{split}
	\label{eq:13}
	\dfrac{\partial}{\partial k_t^2}[f_q^{NS}(x, k_t^2) T_q^{NS}(k_t^2, \mu^2)] + \Theta(k_t^2 - \mu^2)f_a(x, k_t^2)\dfrac{\alpha_s(k_t^2)}{2 \pi k_t^2} \int_0^{z_{max}} z P_{qq}(z) \\
	  = T_q^{NS}(k_t^2, \mu^2) \dfrac{\alpha_s(k_t^2)}{2 \pi k_t^2} \int_x^1 dz P_{qq}(z) f_q^{NS}(\dfrac{x}{z}, k_t^2).
	\end{split}
\end{equation}
This equation is derived in another way in the reference \cite{Guiot_pathology} by redefining the Sudakov form factor as $\tilde{T}_a(k_t^2, \mu^2) = \Theta(\mu^2 - k_t^2)T_a(k_t^2, \mu^2) + \Theta(k_t^2 - \mu^2)$ and inserting this Sudakov form factor inside the \cref{eq:8}. However, the above reference \cite{Guiot_pathology}   claims that with this additional term to the MRW formalism, one can reach to identical UPDFs, both from the I-MRW and D-MRW UPDF, while one can not trace this conclusion \cite{Guiot_pathology}. As it is obvious from what is discussed in this section, in order to obtain this equality, one also needs cutoff-dependent PDFs, in addition of using the \cref{eq:13} instead of the \cref{eq:8} . Therefore, we can expect that none of the prescriptions mentioned in the \cite{Guiot_pathology,Golec_on_KMR} alone, can give us same UPDFs from both the I-MRW and D-MRW UPDFs, and one needs to use them along with each other. In the following section, we provide our numerical results and show such an equivalency.
  \section{Numerical results of the equivalency between I-MRW and D-MRW UPDFs}
\label{sec:3}
In this section, we solve the \cref{eq:9} for the NS distribution with the brute-force method \cite{Brute_force}. For the PDFs at the initial scale we use MSTW-PDF at $1\; GeV$, i.e starting point of this PDFs set, and then evolve PDFs according to the \cref{eq:9}. We perform this evolution for $\mu^2=100\; GeV^2$  and obtain grid files in the $x\textrm{-}k_t^2$ space. Then with the help  of the two dimensional linear interpolation, we can obtain PDFs at different values of $x$ and $k_t^2$. One important point here is that the results with good accuracy can only be obtained, if the grids are dense enough. Now, we are in a position to show our results with the cutoff-dependent NS distribution.

\begin{figure}
	\includegraphics[width=8cm, height=9cm]{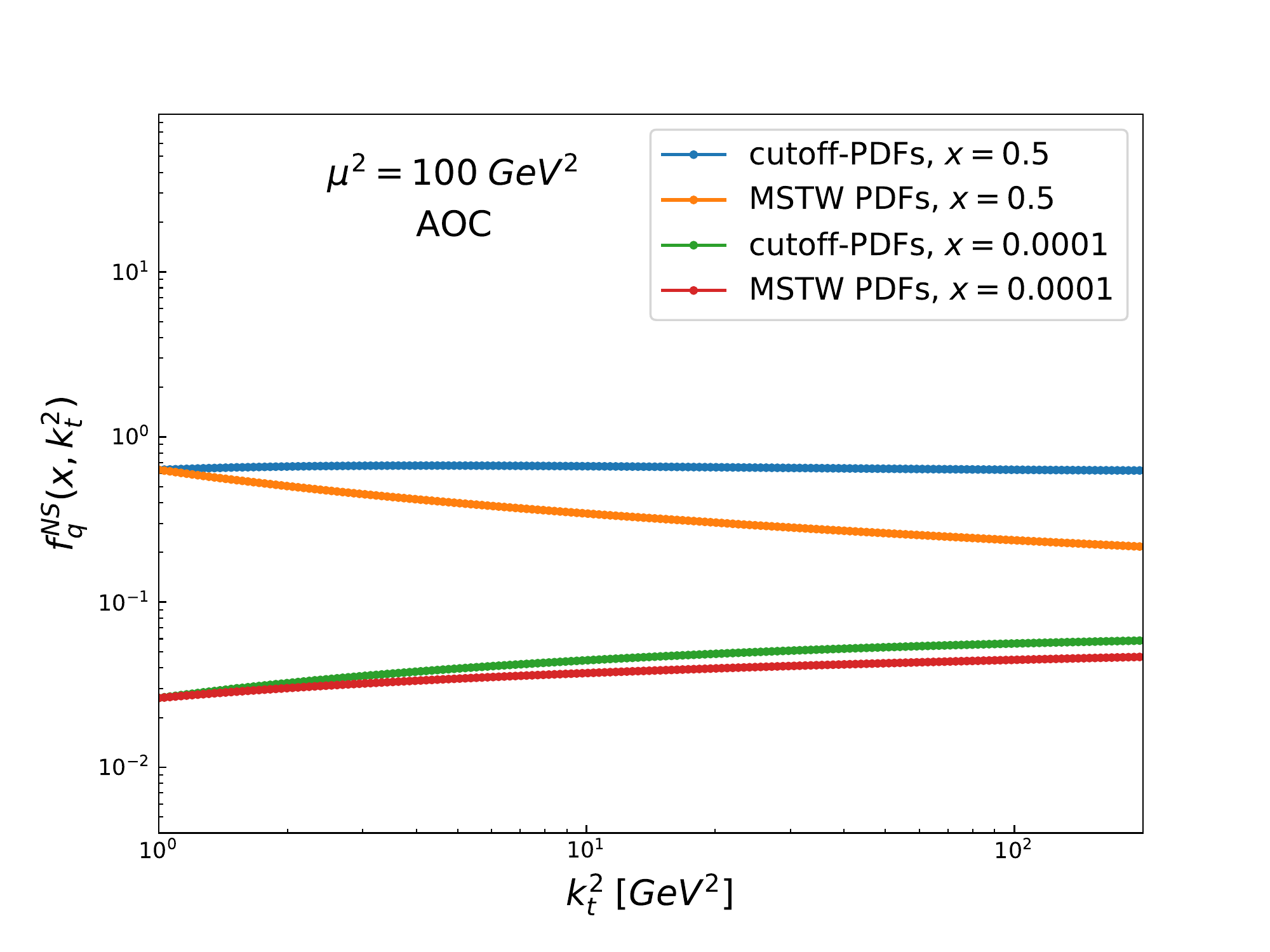}
	\includegraphics[width=8cm, height=9cm]{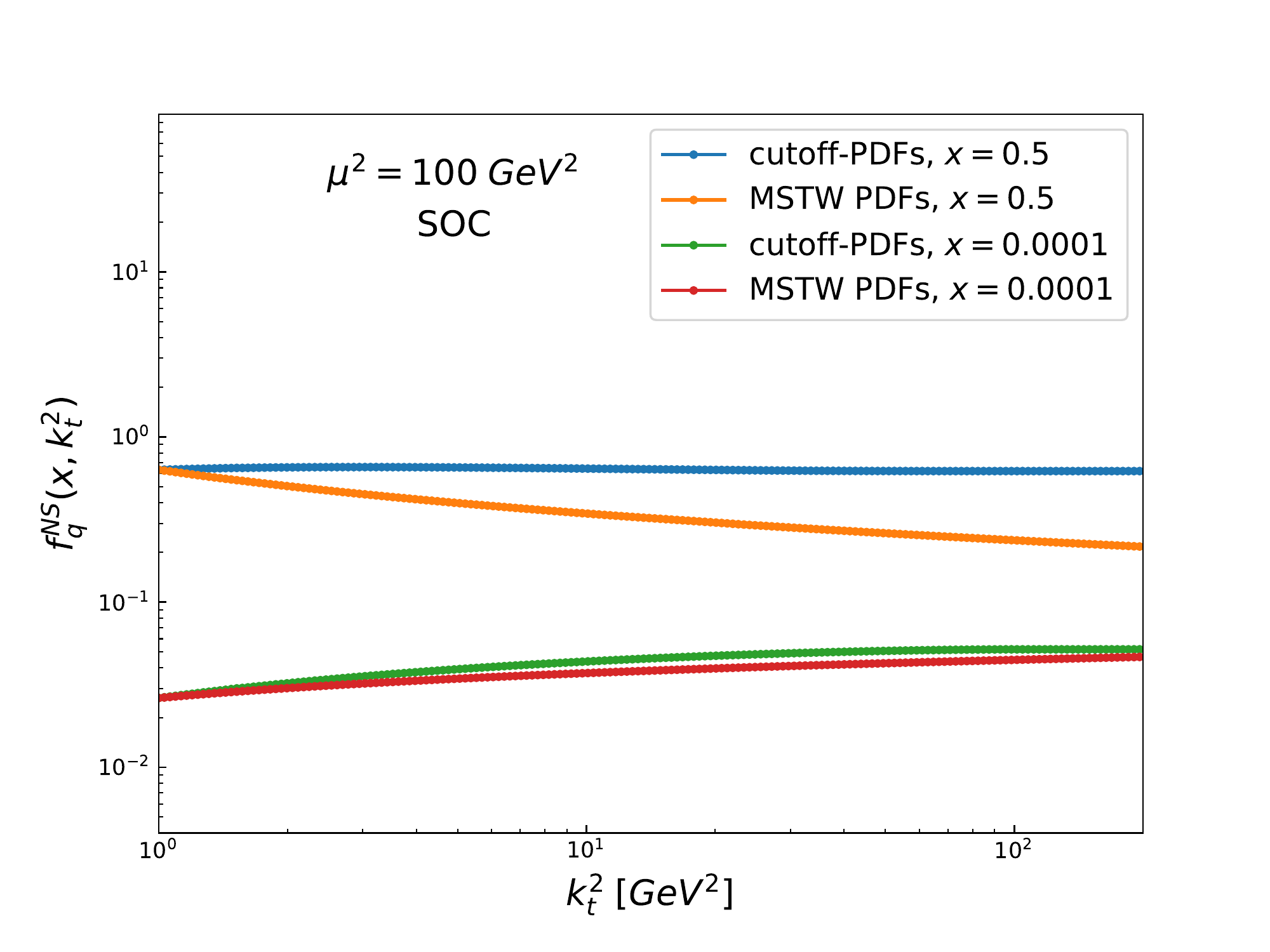}
	\caption
	{
		The left (right) panel shows the comparison of the non-singlet quark distributions of AOC (SOC)   cutoff-dependent PDF with  the MSTW one at $\mu^2=100 \; GeV^2$.
	}
	\label{fig:3}
\end{figure}

\begin{figure}
	\includegraphics[width=7cm, height=9cm]{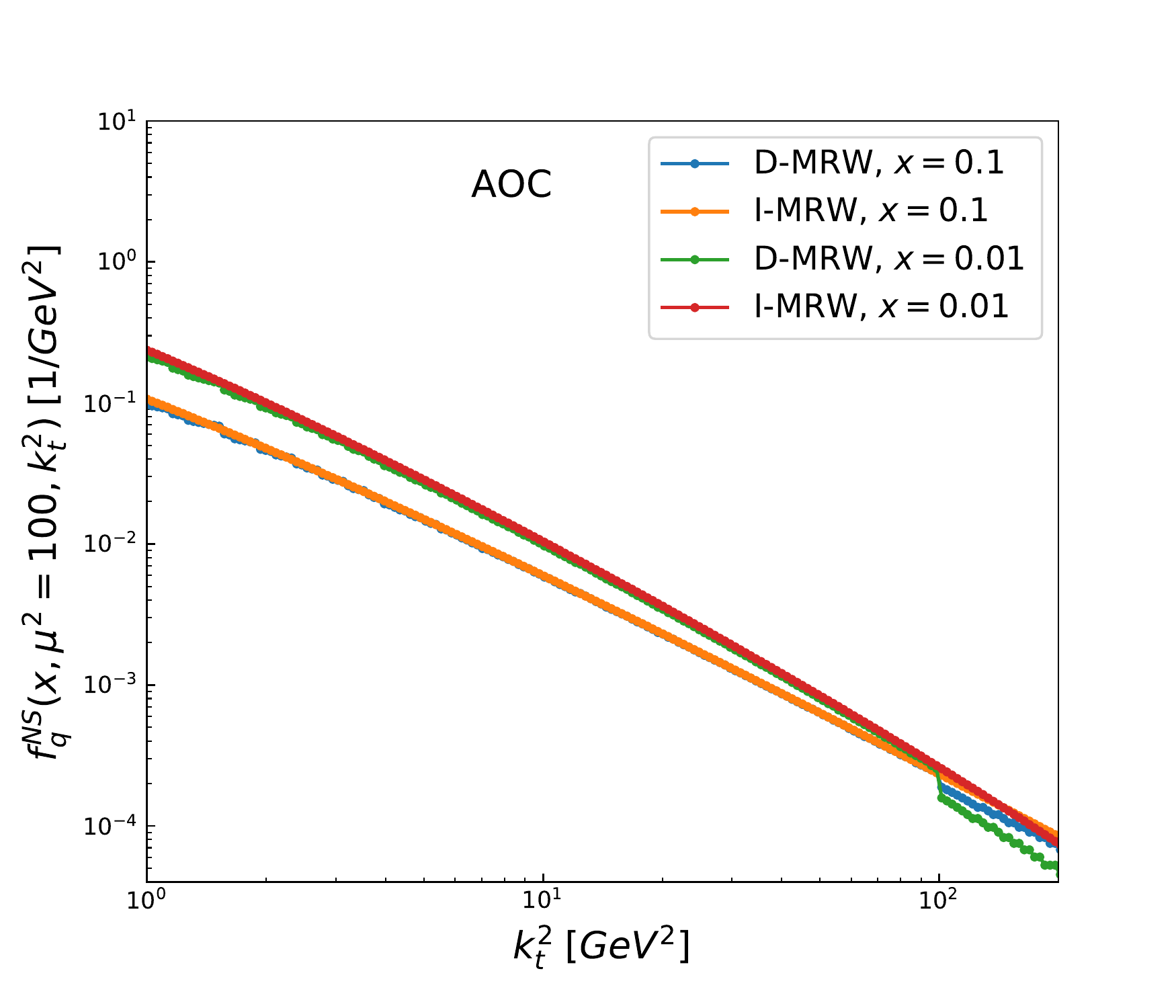}
	\includegraphics[width=7cm, height=9cm]{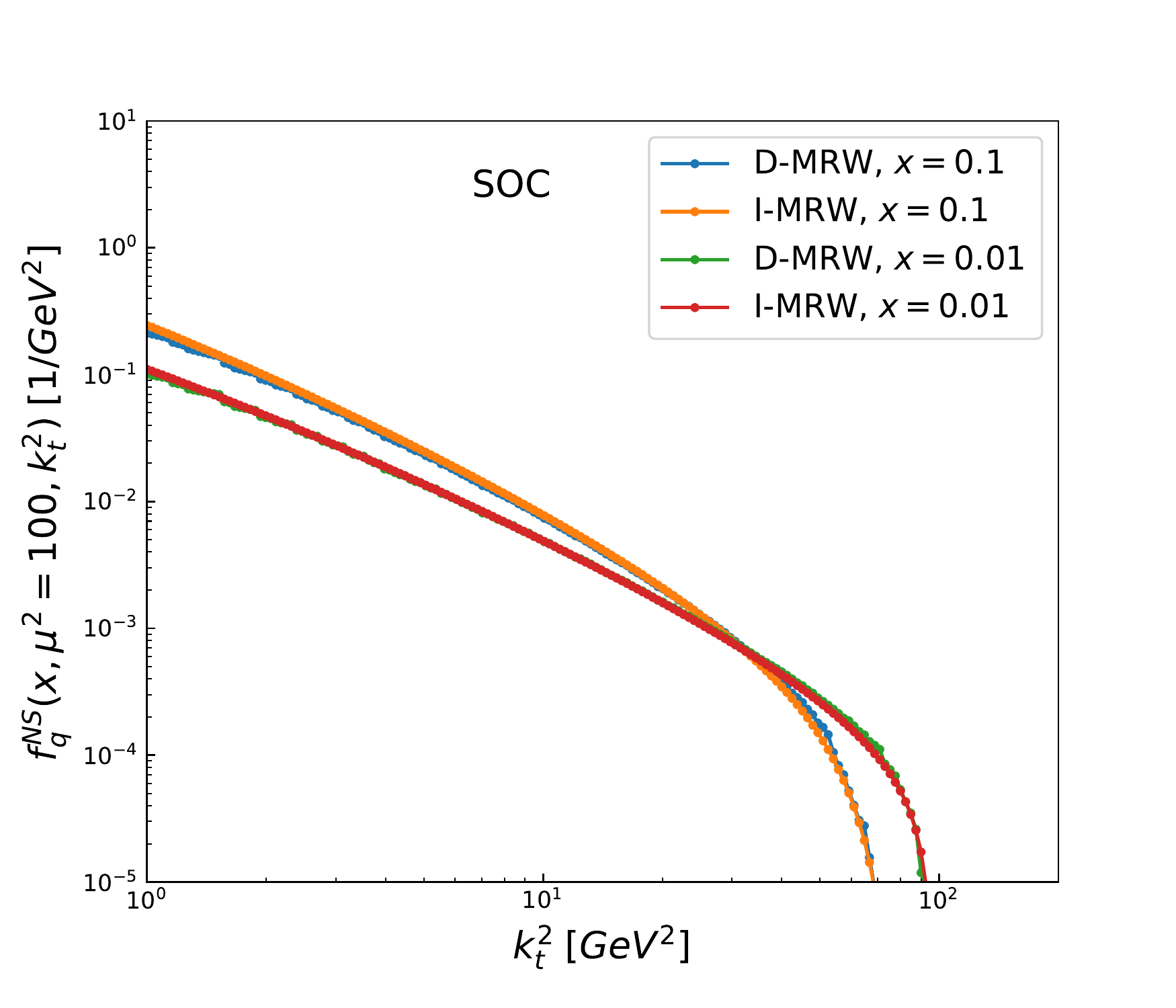}
	\caption
	{
		The left (right) panel  of the \cref{fig:4} shows the numerical result of the \cref{eq:8} with the AOC (SOC)   cutoff-dependent PDF. The differential (integral) UPDF of the \cref{eq:8} is denoted by D-MRW (I-MRW).
	}
	\label{fig:4}
\end{figure}

\begin{figure}
	\includegraphics[width=9cm, height=9cm]{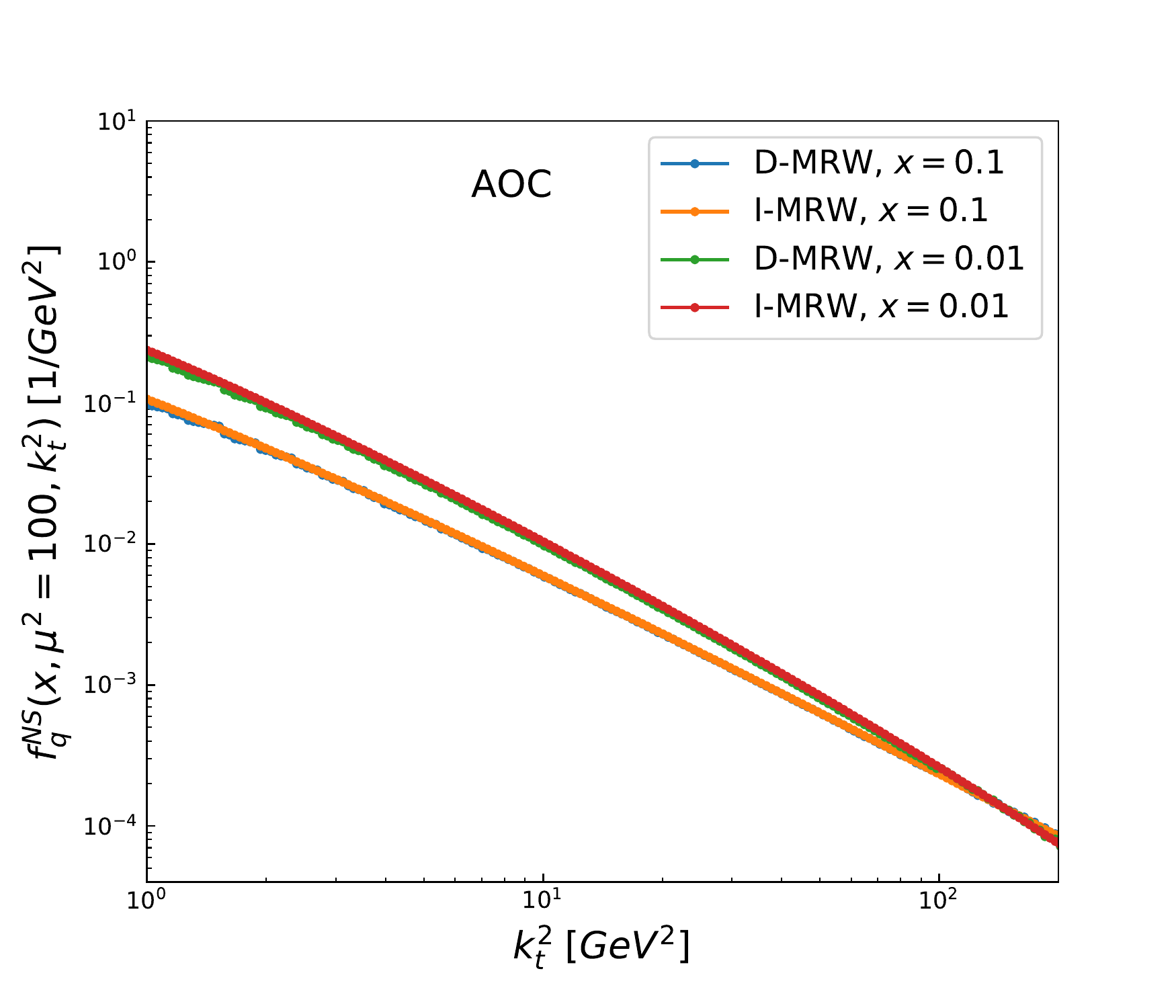}
	\caption
	{
		This figure shows the comparison of the non-singlet quark MRW UPDF employing AOC cutoff-dependent PDF. The differential (integral) UPDFs of the \cref{eq:13} is denoted by D-MRW (I-MRW).
	}
	\label{fig:5}
\end{figure}

First, in the \cref{fig:3} we compare the  cutoff-dependent PDFs, i.e., the AOC and SOC, with the corresponding PDFs of MSTW at $x=0.5$ and $x=0.0001$ to give an insight about their similarities and differences. One can see from this figure that as we approach to the small $x$ and $k_t^2$, i.e., where the choice of the cutoff is not important, cutoff-dependent PDFs and MSTW ones become similar to each other. Another, important point that one can observe in this figure is that, at large $x$, the MSTW-PDF has a decreasing form, while for the cutoff-dependent PDF, such a behavior is not observed. This is the reason that the I-MRW UPDF     with ordinary PDFs are always positive, while if we use the   D-MRW UPDF with these PDFs, it can become negative at large $x$ and $k_t^2$, see D-MRW UPDF at $x=0.1$ in the \cref{fig:1}. However, by using the cutoff dependent-PDFs, one can also obtain, always, positive UPDFs from the   D-MRW UPDF , too. Now, we are in a position to check the claim of the reference \cite{Golec_on_KMR} that with cutoff-dependent PDFs one can obtain equivalency between the I-MRW and D-MRW UPDFs in all $k_t^2$ including the $k_t^2 > \mu^2$.
In the left and right panels of the \cref{fig:4} we show numerical result of the \cref{eq:8} with  AOC and SOC cutoff-dependent PDFs. As can be seen in this figure, the UPDFs obtained with I-MRW and D-MRW UPDFs are the same in the $k_t^2 \leq \mu^2$ region. However, using the \cref{eq:8} leads to the   different I-MRW and D-MRW UPDFs  for the ones with the AOC  cutoff. We should note that if one obtains UPDFs with the SOC, i.e., not using the NS ones, we would   also observe non-equality between the two forms at the $k_t > \mu$.  Henceforth, in the \cref{fig:5}, we show the numerical results of employing the \cref{eq:13} in order to obtain equality between the two forms.  Therefore, it is seen that the cutoff-dependent PDFs alone are not enough for obtaining the same UPDFs both from the I-MRW and D-MRW UPDFs, and using the \cref{eq:13} is essential in obtaining the same UPDFs in all $k_t$ regions. Finally, we should state that, although our results are limited to the non-singlet PDF, but one can generalize them and obtain the same results. 

\section{Conclusions}
\label{sec:4}
In this work, we investigated the equivalency of the differential and integral forms of the MRW UPDF model using angular and strong ordering cutoffs. For simplicity, we only considered the non-singlet quark distribution at the LO level. We first explained the shortcomings associated with the references \cite{Golec_on_KMR,Guiot_pathology}, and then showed that none of the solutions mentioned within these two references are enough to obtain the same UPDFs with the differential and integral forms. Then, we showed that, the methods explained in the aforementioned references are working in certain $k_t^2$ region, and in order to obtain equivalent UPDFs from both the differential and integral forms, one needs to employ both of  these methods, i.e., cutoff-dependent PDFs along with the "modified" differential form.  Finally, employing these two prescriptions, we can obtain unique UPDFs in all $k_t^2$ regions.

\newpage

\end{document}